\documentclass{aastex631}

\shorttitle{Electronic and magnetic properties of silicene monolayer under bi-axial mechanical strain}
\shortauthors{Jafari et al.}
\graphicspath{{./}}

\begin{document}

\title{Electronic and magnetic properties of silicene monolayer under bi-axial mechanical strain: first principles study}

\author[0000-0001-8320-652X]{M.A.Jafari}
\affiliation{Department of Mesoscopic Physics, ISQI, Faculty of Physics, Adam Mickiewicz University, ul. Uniwersytetu Poznańskiego 2, 61-614 Pozna\'n, Poland}

\author[0000-0002-3888-0731]{A.A. Kordbacheh}
\affiliation{Department of Physics, Iran University of Science and Technology, 16846-13114 Tehran, Iran}

\author[0000-0002-1452-7001]{A. Dyrdal}
\affiliation{Department of Mesoscopic Physics, ISQI, Faculty of Physics, Adam Mickiewicz University, ul. Uniwersytetu Poznańskiego 2, 61-614 Pozna\'n, Poland}



\begin{abstract}

Mechanical control of electronic and magnetic properties of 2D Van-der-Waals heterostructures gives new possibilities for further development of spintronics and information-related technologies. Using the density functional theory, we investigate the structural, electronic and magnetic properties of silicene monolayer with substituted Chromium atoms and under a small biaxial strain ($-6\%< \epsilon < 8\%$). Our results indicate that the Cr-doped silicene nanosheets without strain have magnetic metallic, half-metallic or semiconducting properties depending on the type of substitution. We also show that the magnetic moments associated with the monomer and vertical dimer substitutions change very weakly with strain. However, the magnetic moment associated with the horizontal dimer substitution decreases when either compressive or tensile strain is applied to the system. Additionally, we show that the largest semiconductor band-gap is approximately 0.13 eV under zero strain for the vertical Cr-doped silicene. Finally, biaxial compressive strain leads to irregular changes in the magnetic moment for Cr vertical dimer substitution. 

\end{abstract}



\section{Introduction} \label{sec:introduction}
Two-dimensional (2D) crystals are currently of great interest for both applied and fundamental research. The most prominent example of this kind materials is graphene. However, the class of 2D materials is very large and is continuously growing. It contains single-layered materials such as those belonging to the group IV of the periodic table (silicene, germanene and stanene) and to the  group V (arsenene, antimonene, bismuthene), as well as  layered 2D materials like Transition Metal Dichalcogenides and MXenes.
Among them silicene seems to be a promising material due to its  compatibility with existing silicon-based electronic devices. Because of  sp3 hybridization, silicene is the 2D buckled hexagonal  lattice of silicon atoms, and  is considered as a one of alternative materials to graphene \cite{2,3,4,5,6}. Research activities on silicene significantly increased after its successful synthesis under UHV conditions on several substrates,  like for instance on Ag(111),  ZrB2(0001), MoS2(0001), and Ir(111) \cite{7,8,9,10,11}.
Silicene displays several interesting characteristics, which have been revealed by recent experimental and theoretical investigations. These include for instance: (i) a remarkable spin–orbit coupling parameter, that leads to the energy gap of 1.5 meV \cite{16,17,18}  at the Dirac point, which is much larger than that in graphene (24 $\mu$eV) \cite{3,19,21}; (ii) electrically tunable bandgap; (iii) the phase transition from a spin Hall topological insulator to a band insulator~\cite{31}; (iv) the strain-induced tunable bandgap \cite{22,23,24}, and (v) promising electric and thermoelectric characteristics ~\cite{33,34,36,85,86,87,88}.  The energy gap in silicene makes it promising for applications,  however the realization of stabile monolayer of silicene  is still problematic \cite{1}. For example, to achieve high on-to-off current ratios and a perfect switching capability, the material exploited for Field-Effect-Transistors (FETs)  is usually required to have a fairly large bandgap \cite{11}, significantly  larger than that mentioned above for silicene.

Electronic and magnetic properties of silicene can be tuned by impurities, i.e., magnetic atoms built into the monolayer structure.
Importantly, recent achievements of nanotechnology enable precise arrangements of impurities (including regular lattices), so by doping one can modify electronic properties in a controllable way.
The single-side adsorption of alkali metal atoms on silicene has been reported to give rise to a bandgap of approximately 0.5 eV \cite{25,26}, that is much larger than that due to intrinsic spin-orbit interaction. If both sides of the single-layer silicene have been saturated with hydrogen atoms, the formed composite system has been shown to be a kind of a nonmagnetic semiconductor \cite{27}. Based on the previous studies, the ferromagnetic characteristics of silicene can be attributed to the single-side hydrogenation \cite{28}.

\begin{figure}[t]
\centerline{\includegraphics[width=0.8\columnwidth]{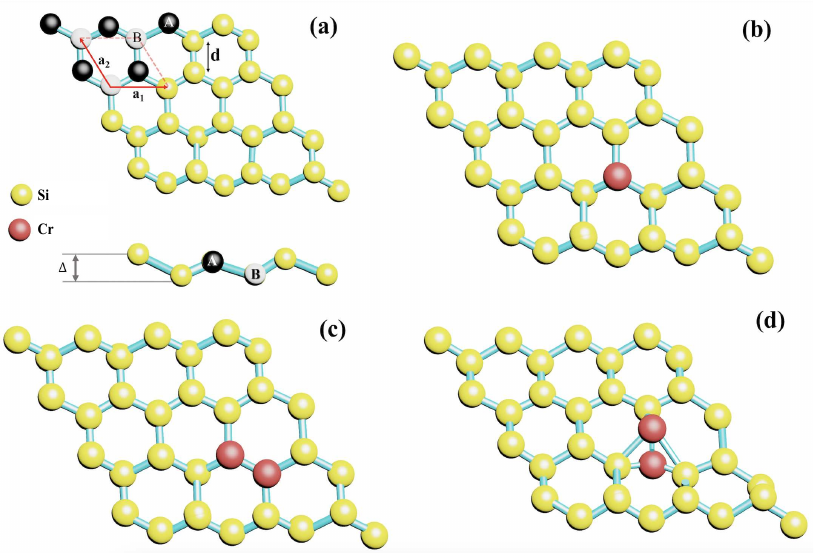}}
\caption{(a) Hexagonal crystal structure of silicene, with $d$ being the atom spacing and  $\Delta$ denoting the buckling parameter. The unit cell is indicated by he red dashed lines, and the basis consists of two Si atoms, labelled with A and B. (b-d) Types of impurity substitutions: monomer (b), horizontal dimer (c), and vertical dimmer (d). }
\label{fig:fig1}
\end{figure}

Strain engineering enables external control of electronic characteristics of semiconductor heterostructures and nanomaterials. This technique is widely applied to nano-electro-mechanical and nano-opto-mechanical systems, as well as to MOSFETs \cite{29,30}.
In turn, the in-plane strain  in silicene  leads to a modification of the electronic structure and its transport characteristics \cite{66,32}.

An important question is whether a significant gap in the spectrum of silicene can be induced by the strain, especially at low values of strain. If this is the case, tunable silicene devices would be of great practical importance. According to earlier density functional theory (DFT) calculations, a gap in the spectrum of silicene can be open under arbitrary uniaxial strain. Its magnitude varies non-monotonically with the strain. These findings were  supported by other ab-initio (without empirical parameters and from first principles) calculation \cite{68}. For similar strain magnitudes, however, the two gap calculations were not in agreement. Recently, Pereira et al. \cite{39} have questioned the accuracy of those conclusions by applying the Tight Binding (TB) approach. According to their findings, a spectral gap can only be achieved for +20\% uniaxial deformations. Furthermore, this effect  highly depends on the deformation route regarding the rudimentary lattice. The aforementioned findings are in accordance with the studies of Hasegawa et al.  \cite{40}, indicating that there is a robust gapless Dirac spectrum with regard to arbitrary and not extremely large changes in the nearest neighbour hopping parameters. Moreover, employing the TB model, Wunsch et al. \cite{41} found  that the semi-metallic phase appears for hopping parameter expansion.

The results of ab-initio calculations \cite{42} are in agreement with the gapless situation presented in \cite{43}.
The inconsistency between various ab-initio calculations ware partially related to the fact that due to strain the Dirac points shift from the high symmetry points of the Brillouin zone. This resulted in arriving at the wrong conclusion that a bandgap is achievable for any strain.
In addition, Faccio et al. also performed  DFT research and calculated the impact of $ \approx 2\% $ strain in silicene nanoribbon [43].
Keeping in mind that mechanical strain can substantially change the physical properties of silicene, we have performed  detailed ab-initio calculations of the magnetic and electronic characteristics of silicene with inserted dimers and monomers of Cr atoms.
Then, we have analysed  the influence of a biaxial strain for up to $ \approx \pm 8\% $ deformations.
We have shown that magnetic and electronic characteristics of silicene with substitutional impurity atoms can be easily controlled by various kinds of strains (i.e., substrate-induced strains or external mechanical forces).
It is worth to note, that the gap due to strain and doping is significantly larger than the gap induced by the intrinsic spin-orbit interaction in undoped and unstrained silicene.

\section{Methodology and strain-dependent structural properties} \label{sec:methodology}
To study electronic and magnetic properties of silicene under mechanical strain we have used the DFT  method \cite{44} within the Perdew-Burke-Ernzerhof (PBE) generalized gradient approximation (GGA) \cite{45}  form of the exchange-correlation functional, as implemented in the
QuantumATK code package ver.~S-2021.06~\cite{45-1,45-2}. The PseudoDojo collection of optimized norm-conserving Vanderbilt (ONCV) pseudopotentials and ultra-basis set have been used for the optimization of structures and for further  calculations  \cite{46}. For the Brillouin zone integration we have taken 8$\times$8$\times$1 Monkhorst-Pack k-points in self-consisting calculation (SCF), and the mesh cut-off of energy has been set to 450 Ry. Structures ware relaxed until the forces on each atom were less than 0.05 eV/$\AA$ and relative convergence for the Self-Consistent Field (SCF) energy is reached until $10^{-5}$ eV/$\AA$. We have used 15 $\AA$ vacuum region to prevent interaction of two neighboring layers along the $c$-axis \cite{47,48,49}.

The atomic structure of silicene should be characterized before modelling the electronic structure \cite{50}. The first  structure optimization process for a single layer of Si was reported  by Takeda and Shiraishi \cite{51}. By analogy to graphite, they defined a hexagonal lattice for Si atoms (with a periodicity perpendicular to the plane with a large vacuum layer of minimally 10 $\AA$), and then varied the in-plane lattice constant and positions of the basis atoms (Figure 1(a)) inside the unit cell, while keeping constant the imposed D3d symmetry \cite{52}. According to their results, the buckled structure has a lower total energy, with a local minimum for $a = 3.855$ $\AA$ and a deformation angle of 9.9$^{\circ}$, when compared to the energy of a flat structure \cite{50}.

In the present study, the  4$\times$4 monolayer of silicene was characterized systematically (thus the unit cells were repeated up to four times in the $x$ and $y$ directions) using DFT \cite{53}.
The Si atom bounded to   three nearest neighbour surrounding Si atoms with the Si-Si bond length of $2.28\,\AA$ and lattice parameter $a$ of $3.84\,\AA$  was assumed prior to structural relaxation.
The optimized lattice parameter, $a = 3.86\,\AA$, correlates quite well with other data, even though the standard GGA functional method  overestimates it. The buckling parameter $\Delta$ is $0.46\,\AA$, which is also consistent with other studies \cite{55,59,62,63,64}.

In this paper we analyse three different types of substitution in silicene monolayer by Chromium adatoms as presented in Figure~\ref{fig:fig1}. These are: (i)  monomer substitution, where one Si atom in the supercell is substituted by  Cr atom  ($3.2\% $ substitution), as presented in Figure~\ref{fig:fig1} (b);  (ii) horizontal dimer (HDimer) substitution with two neighbouring Si atoms substituted by Cr atoms ($6.25\%$ substitution), see Figure~\ref{fig:fig1} (c); (iii) vertical dimer (VDimer) substitution, where one Si atom is replaced by two Cr atoms ($3.2\%$ substitution), as shown in Figure~\ref{fig:fig1} (d). These three structures were modeled within the optimized 4$\times$4 supercell geometries, and we analysed behaviour of the electronic and magnetic properties with the strain.

\begin{figure}[t]
\centerline{\includegraphics[width=0.65\columnwidth]{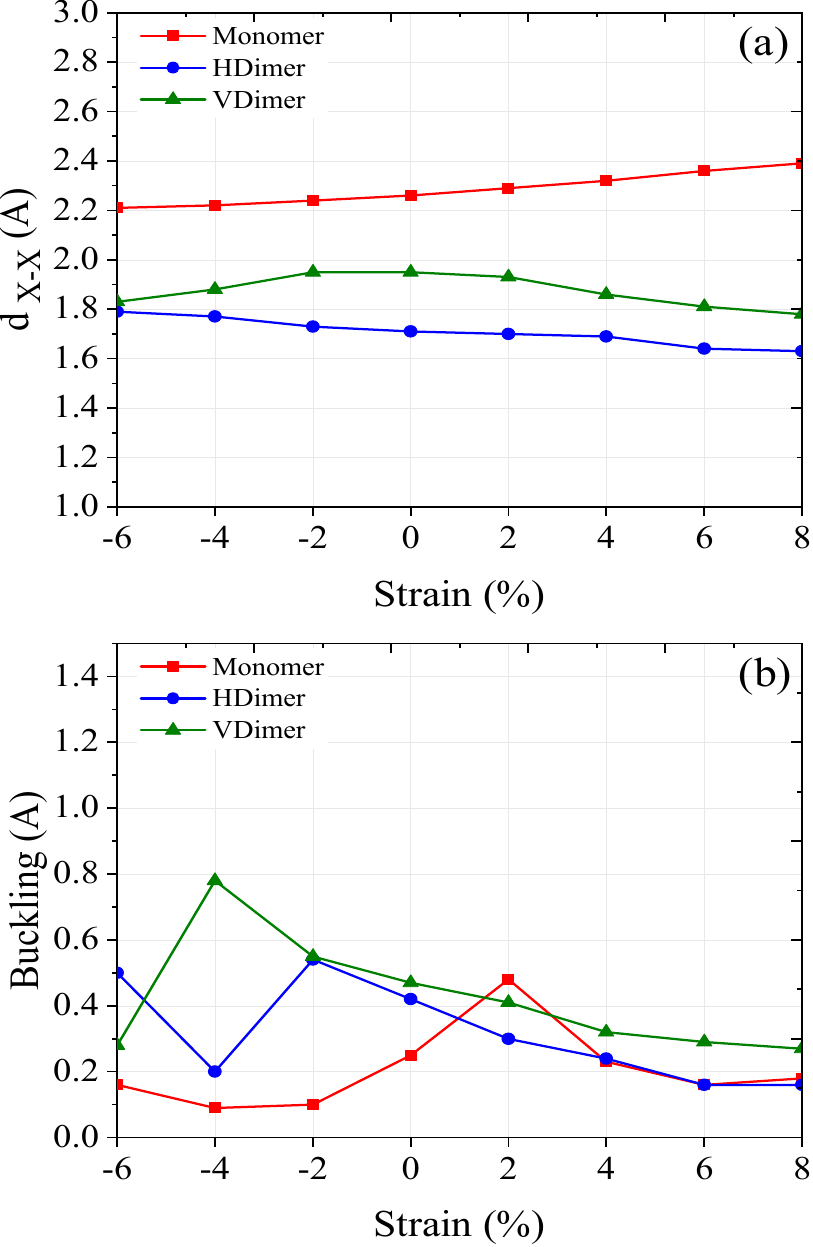}}
\caption{Variation of the Cr-Si  and Cr-Cr bond lengths, $d_{X-X}$ (Fig.2.(a)), as well as of the buckling parameter (b) under the biaxial tensile and compressive strain for the  monomer, horizontal dimer,  and vertical dimer substitutions.}
\label{fig:fig2}
\end{figure}

The strain is defined as a deformation resulting from external loads or forces, that may be calculated by the following equation:
${\epsilon={\Delta a} /a_0}$ with ${\Delta a = a - a_{0}}$, where $a$ is the lattice parameter of the strained silicene and ${a_0=3.86\,\AA}$ is the lattice constant of the unstrained silicene \cite{66,68,67}. The model of strained unit cell for the tensile and compression strains is achieved by varying the lattice constant along the lattice vectors by the following substitution $a \rightarrow \epsilon a$.
Accordingly, the uniaxial tensile strain or compression ($\Delta a \gtrless 0$ respectively) is realized by modification of the lattice constant in the
 $x$ direction, that is by modification of the lattice vector $\mathbf{a}_{1}$ (mechanical force applied along the lattice vector $\mathbf{a}_{1}$), whereas the biaxial tensile or compressive strain is described by modification of  both $\mathbf{a}_{1}$ and $\mathbf{a}_{2}$ lattice vectors (forces are oriented along both lattice vectors).

The presence of biaxial strain affects the buckling parameter, $\Delta$, as well as the Cr-Si and Cr-Cr bonds  in the silicene monolayer with monomer and dimer substitutions, respectively.
Figure \ref{fig:fig2} presents the basic parameters  of the relaxed structures as a function of strain and also for all the substitutions under consideration.

For the monomer substitution (see  Figures \ref{fig:fig2}(a)), the bond length between Cr and Si atoms, $d_{\mathrm{Cr-Si}}$, slightly increases with the strain. However, there is no clear universal behavior of the buckling parameter with strain, though one can see that this parameter reaches a maximum for a specific strain  equal $2\%$, and then decreases with increasing magnitude of either tensile or compressive strain.

For the silicene with HDimer substitution, the tensile strain reduces monotonically the bond length $d_{\mathrm{Cr-Cr}}$ and the buckling parameter. However, for the compressive strain one observes increase of the bond length for Cr-Cr dimers. In turn,  the buckling parameter for Cr-Cr HDimers  varies nonmonotonously with the magnitude of compressive strain, i.e., it reaches a minimum at some magnitude of the strain, see Figure \ref{fig:fig2}(b).

\begin{figure*}[t]
\centerline{\includegraphics[width=1.0\columnwidth]{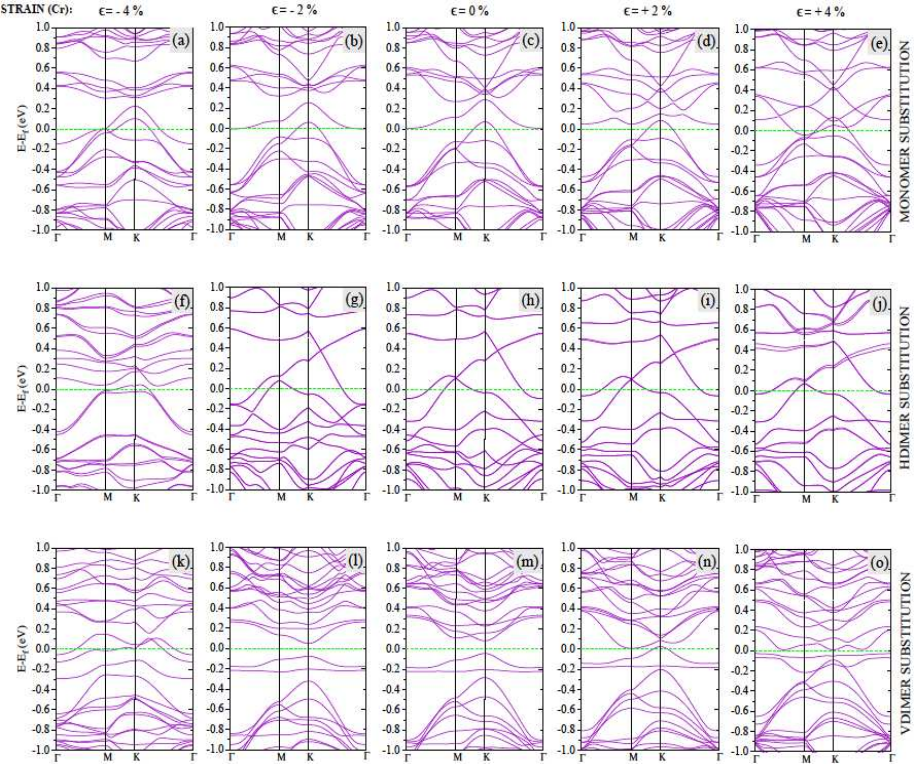}}
\caption{The electronic energy spectrum of silicene under selected values of biaxial strain (from $-4\%$ to $+4\%$) in the presence of spin-orbit interaction, and for Cr doping with the three types of substitution: (a-e) monomer, (f-j) HDimer, and (k-o) VDimer. }
\label{fig:fig3}
\end{figure*}
In turn, for the VDimer substitution, the strain only very slightly affects the bound lengths reducing it when the strain takes absolute values larger  than 2$\%$.
The corresponding buckling parameter decreases with increasing tensile strain and increases with increasing magnitude of the compressive strain up to $4\%$, where it reaches a maximal value, and then decreases with a further increase of the magnitude of compressive strain.

At this point, one should note that high strain values are experimentally difficult to obtain. \cite{68,69,70,71,72,73}.
The recent reports about graphene-like monoatomic crystals indicate that the strain around $4\%$ can be relatively easily obtained.
Accordingly, in this paper we have restricted our considerations to the strain ranging from $-6\%$ to $8\%$.

\section{Strain-dependent electronic and magnetic properties} \label{sec:transport}

It is well known that strain has a significant impact on electronic and magnetic properties of 2D crystals \cite{74,76,77,89,90}.
One of the consequences of the strain in 2D crystal is the electronic bandgap engineering, i.e., strain-induced bandgap opening \cite{33} or a direct–indirect–direct bandgap transition in green phosphorene \cite{79,80}.

Figure \ref{fig:fig3}  presents the band structures of strained and unstrained silicene doped with Cr for the three different types of substitution (Monomer, HDimer and VDimer), as discussed in Section 2. The band structure has been calculated along the high-symmetry points of the Brillouin zone, i.e., along the $\Gamma$-M–K–$\Gamma$ path.
In turn, Table 1 collects information about the band gap in silicene under strain and with different types of Cr substitution. The electronic band structure of undoped silicene monolayer under  strain is presented in \ref{AA}. Here one needs to remind, that strain in undoped silicene does not open a gap. A small gap appears only due to spin-orbit interaction. When neglecting this interaction, the gap in undoped silicene remains equal to zero  (at the Dirac points), see also Figure \ref{fig:fig5} in Appendix A.
Apart from this, due to hybridization of the 3d-impurity states and those of pure silicene, the band structure becomes remarkably modified by doping. Each  state of the doped system includes in general contributions due to 3d-transition metals  as well as due to silicon atoms. To show this explicitly, we have presented the so-called {\it fat-bands} structure, where the two contributions are indicated explicitly  with different colors. In  \ref{AB} we present {\it fat-bands}  calculated for silicene monolayer doped with Cr-atoms (i.e., the band structure projected over orbitals of Silicone and  Chromium atoms). The corresponding  results are shown in Figure \ref{fig:fig6}, and from this figure one can estimate whether  a particular band contributes to conductivity or not. If it is due to impurities only and is dispersionless around  the Fermi level, it does not contribute to  conductivity. If however it is dispersive around  the Fermi level, then even though the silicon contribution is small, it contributes to conductivity. These features have been taken into account when determining the  band gaps. In \ref{AB} we show the {\it fat-bands} for silicene with  Cr-Monomers, Cr-HDimers  and Cr-VDimers.

The silicene with Cr-monomers is either metallic or semimetallic,  with the Fermi energy crossing the valence bands. In the presence of compressive strain the energy gap becomes opened in the spectrum above the Fermi energy, while the Fermi level is still inside the valence bands. In the presence of tensile strain the system moves from semimetallic to metallic one.
In turn, for silicene with Cr-HDimer substitution, the electronic structure is only slightly affected by the strain.
The most promising effect of strain on the electronic structure can be observed for silicene monolayer with Cr-VDimer substitution.  Without strain the system is a semiconductor with the energy gap equal 0.13 eV.  Applying compressive strain one can close the energy gap and move the Fermi level  to the valence bands. The tensile strain, in turn,  leads to reduction of the energy gap for the 4$\%$ strain, and leads to its complete closure for strain $\ge 6$ $\%$.

\begin{figure*}[t]
\centerline{\includegraphics[width=1\columnwidth]{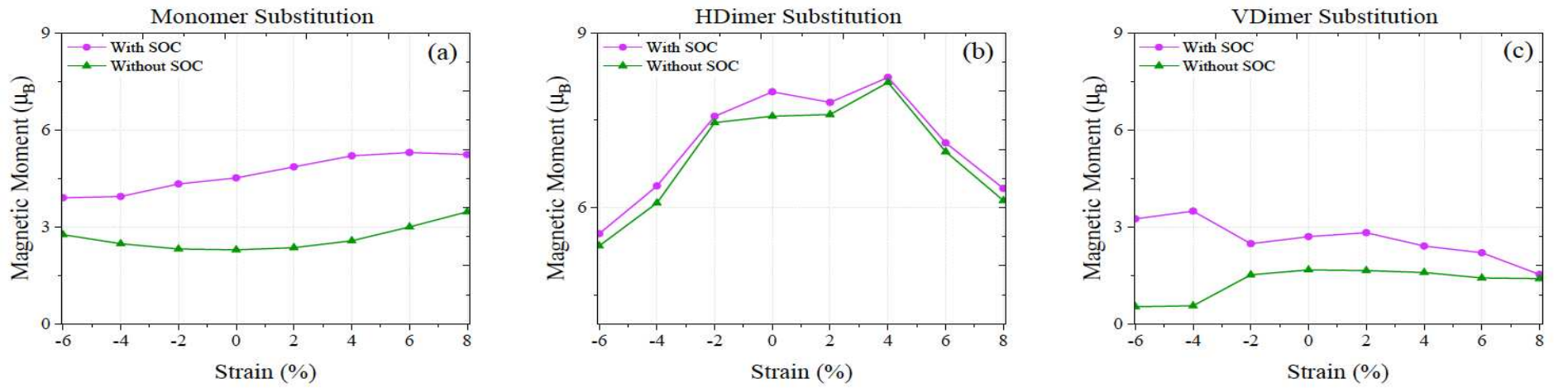}}
\caption{Magnetic moment behaviour as a function of the applied biaxial strain for the Cr substitution obtained based on calculations with and without spin-orbit coupling (SOC). The green curves (no SOC) correspond to the collinear calculations discussed in \ref{AC}.}
\label{fig:fig4}
\end{figure*}

\begin{table}[]
\caption{Band Gap (~eV) changing for Cr substitution during the biaxial strain and applied spin-orbit interaction}
\begin{tabular}{|c|c|c|c|c|c|c|c|c|}
\hline
Strain (\%) & -6\% & -4\% & -2\% & 0\%  & 2\% & 4\%  & 6\% & 8\% \\ \hline
Monomer-Cr  & 0    & 0    & 0    & 0    & 0   & 0    & 0   & 0   \\ \hline
HDimer-Cr   & 0.1  & 0    & 0    & 0    & 0   & 0    & 0   & 0   \\ \hline
VDimer-Cr   & 0    & 0    & 0.12 & 0.13 & 0   & 0.05 & 0   & 0   \\ \hline
\end{tabular}
\end{table}

Our calculations also reveal the impact of biaxial strain on the magnetic characteristics of the silicene monolayer doped with the specified above Monomers, HDimers and VDimers of Cr atoms.
The corresponding results are presented in
Figure \ref{fig:fig4}, where
spin polarization of silicene monolayer doped with Cr atoms is shown as a function of strain (the three types of substitution are presented).
Situations in the presence of spin-orbit interaction and that in the absence of spin-orbit coupling are shown there.
In the latter case, the calculation procedure is described in \ref{AC}.
For silicene with the Cr-monomers, the magnetic moment varies monotonously with the strain when the spin-orbit coupling is included, while in its absence, the magnetic moment increases with the magnitude of strain, both tensile and compressive. Moreover, the difference between these two situations is relatively large.
In turn, for the case of Cr-HDimer, the magnetic moment decreases with the compressive and tensile strain, and the difference between the cases with and without spin-orbit interaction is small. For the silicene monolayer doped with Cr-VDimers, the magnetic moment only weakly depends on the strain for strain larger than $-4\%$, and the difference between the case with and without spin-orbit interaction is small except the strain below $-4\%$, where this difference is relatively large.
It is worth noting that the largest magnetic moment for the unstrained system is for silicene monolayer with Cr-HDimers, where the changes in the magnetic moment due to strain are also most pronounced.

\section{Conclusions} \label{sec:summary}

In this work we presented detailed study of electronic and magnetic properties of silicene doped with Cr atoms in one of the three doping schemes, i.e.,  monomer, HDimer, and VDimer substitutions. Numerical results based on the DFT calculations clearly show that the way of substitution may substantially change the structural, electronic and magnetic behaviour of the silicene under strain. The interplay of doping and strain may be used to engineer band gap, and thus also character of transport properties from metallic to half-metallic or semiconducting.
Such a strain-induced engineering of transport properties is  important from the practical point of view as it may be used in various spintronics and/or logic devices.  It is expected, that controlling current and magnetic state with a strain opens new route for nanoelectronics of future generation.
\\
\section{Acknowledgments}
We thank Prof. J. Barnaś for valuable discussions and reading this manuscript.
This work has been supported by the Norwegian Financial Mechanism 2014-
2021 under the Polish-Norwegian Research Project NCN GRIEG “2Dtronics”
no. 2019/34/H/ST3/00515.

\begin{figure*}[h]
\centerline{\includegraphics[width=0.7\columnwidth]{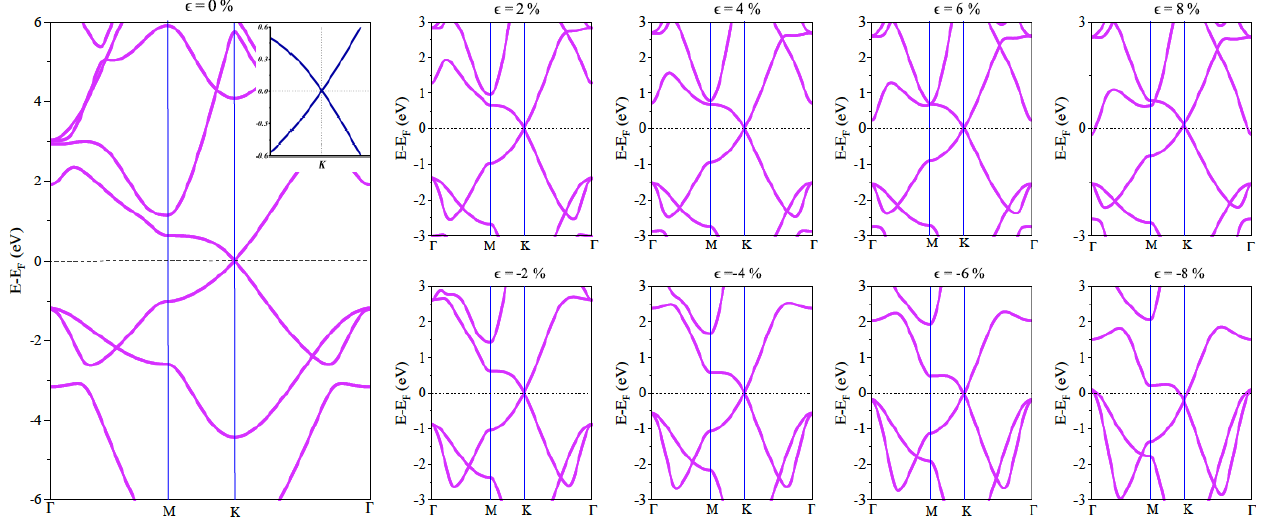}}
\caption{The electronic band structure of undoped silicene monolayer under biaxial strain from $-8\%$ to $+8\%$ and in the absence of spin-orbit interaction.}
\label{fig:fig5}
\end{figure*}

\appendix

\section{The electronic band structure of undoped silicene}
\label{AA}

To understand the interplay of the effects due to strain and doping, it is advisory to analyse the impact of strain on the undoped silicene. In Figure \ref{fig:fig5}  we present the electronic structure of pure silicene monolayer subject to a biaxial strain from $-8\%$ to $+8\%$, and  in the absence of spin-orbit coupling. It is evident, that the strain has a significant a impact on the band structure. However, the band gap remains zero as in the absence of strain. The valence and conduction bands tough each other at the Dirac K points, independently of the strain, which preserves semi-metallic behavior of the system with increasing strain, either tensile or compressive. The strain does not open a gap, however, it modifies asymmetry of the Dirac cones.

\section{The fat-band structures}
\label{AB}
\begin{figure*}[t]
\centerline{\includegraphics[width=1\columnwidth]{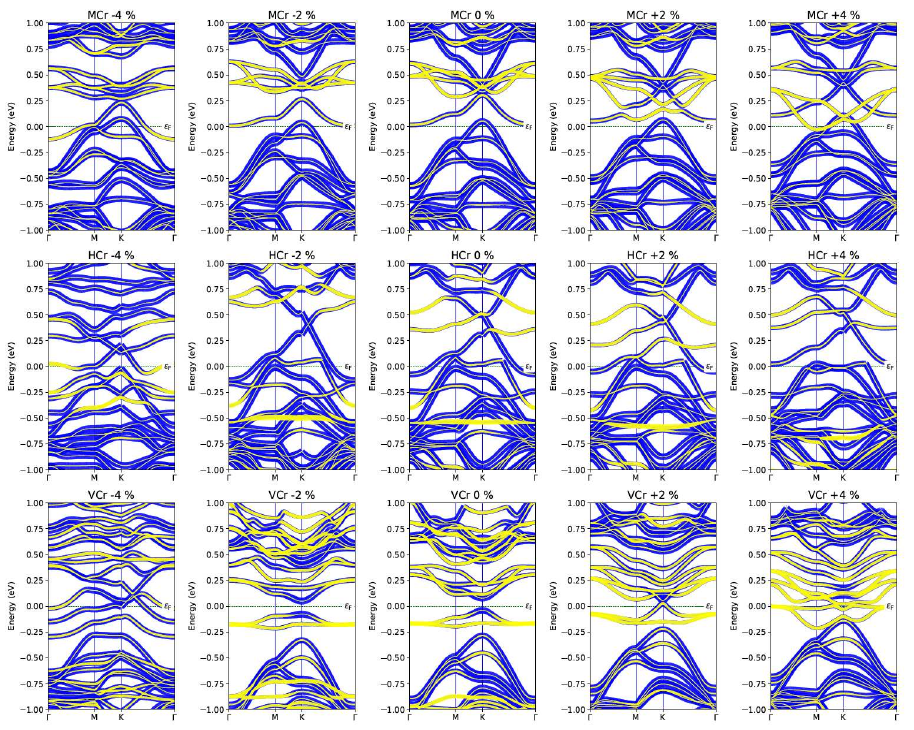}}
\caption{A fat-band representation is used to indicate the projected weights in a silicene monolayer with substituted Chromium atoms in the (top) monomer, (middle) HDimer, and (bottom) VDimer forms. The yellow and blue colors represent the transition metal and silicon atoms contributions, respectively.
}
\label{fig:fig6}
\end{figure*}

As already mentioned above, due to hybridization of the 3d-impurity states and those of pure silicene, the band structure becomes remarkably modified by doping. Each  state of the doped system includes then, in general, contributions from the 3d-transition metals  as well as due to the silicon atoms. We show this explicitly in Figure \ref{fig:fig6}, where the two contributions are indicated  with different colors  for silicene monolayer doped with Cr-atoms. We show there the fat-bands for Cr-Monomers, Cr-HDimers  and Cr-VDimers. From this figure one can evaluate whether  a particular band contributes to transport or not, and this was  taken into account when determining the band gaps presented in Table 1.

\section{The electronic band structure of silicene doped by Cr atoms from collinear calculations}
\label{AC}

\begin{figure*}[t]
\centerline{\includegraphics[width=1.05\columnwidth]{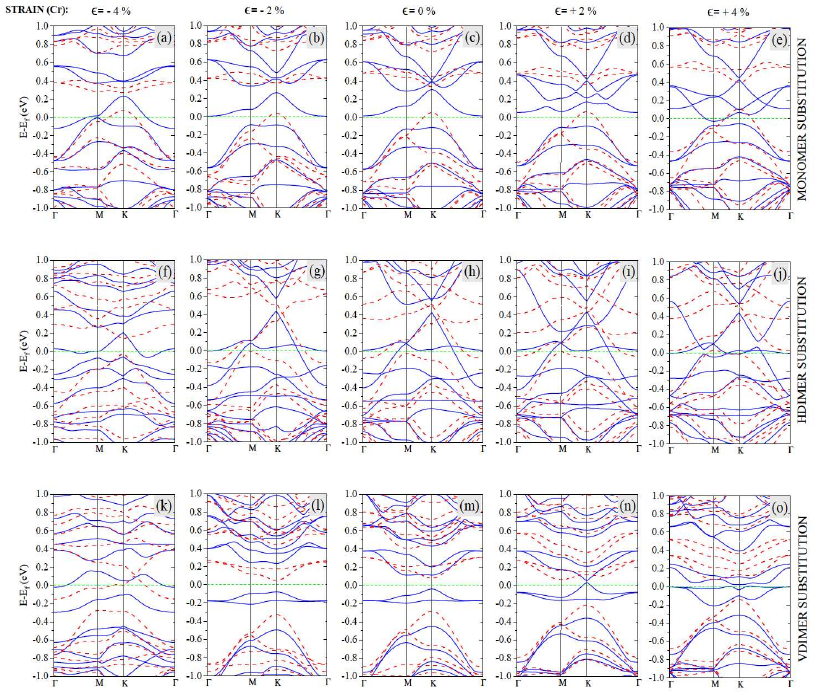}}
\caption{The electronic energy spectrum of silicene under selected values of biaxial strain, from $-4\%$ to $+4\%$,  for Cr doping in case of three types of substitution: (a-e) monomer, (f-j) HDimer, and (k-o) VDimer. The solid (blue)  and dashed (red)  lines correspond to the  spin-up and spin-down, respectively.}
\label{fig:fig7}
\end{figure*}

In this section we present electronic spectrum of silicene monolayer doped by Cr atoms in the presence of biaxial strain. As before we consider the three types of substitutions discussed in the main text. The electronic structure has been obtained from collinear (spin polarized) calculations. This procedure is applicable in the absence of spin-orbit interaction, where spin-up and spin-down states are well defined.   Figure~\ref{fig:fig7} presents the corresponding  electronic band dispersions, where the solid blue and dashed red lines correspond to the spin-up and spin-down states, respectively.

\bibliography{Ref-strain}{}
\bibliographystyle{aasjournal}


\end{document}